# KiWi - OCC Key-Value Map

Assaf Yifrach
Tel-Aviv University
assafyifrach@mail.tau.ac.il

Niv Gabso
Tel-Aviv University
gabsoniv@gmail.com

## Abstract

We take a relatively fresh wait-free, concurrent sorted map called KiWi, fix and enhance it. First, we test its linearizability by fuzzing and applying Wing&Gong [2] linearizability test. After fixing a few bugs in the algorithm design and its implementation, we enhance it. We design, implement and test two new linearizable operations sizeLowerBound() and sizeUpperBound(). We further compose these operations to create more useful operations. Last, we evaluate the map performance because previous evaluations became obsolete due to our bug corrections.

# 1. Preface

## 1.2 KiWi data structure overview - emphasis on intra-chunk management.

KiWi is a relatively new key-value concurrent balanced sorted map by Sulamy et. al [1]. It supports linearizable get(key), put(key, val), scan(minKey, maxKey). All its operations are lock-free, while get() and scan() are also wait-free. The data structure synchronization is based primarily upon CAS operations and atomic increments. It applies common wait-free synchronization techniques: try and RETRY (OCC), version control and helping.
We will not delve into all the details of its design, only cover the overall design and specific details that are critical for understanding our work. For more details about KiWi, read [1]. However, note that in [1], the pseudo code sections and their explanations contain some coarse mistakes. So for a correct and whole understanding, one must read the code.

### 1.2.1 Data structure organization

KiWi organizes the data in a sorted linked list of chunks. Each chunk $c$ consists of all entries in the range $[c.minKey, c.next.minKey)$.
**Variables**
**GV**: Global version - atomic integer. Incremented once by every scan operation.
**INDEX**: ConcurrentSkipList<Integer, Chunk> - index for the various chunks, for fast find of a chunk according to an index.
**PSA** - Pending scans array - array for ongoing scans. Used for safe rebalancing, we will not discuss it further.
Each **Chunk** consists of:
- **PPA** - An array of pending put operations. Each thread has a cell in the PPA. Each thread may edit only its cell in the PPA.
- **Ordered Array** - A sorted linked list of map elements, implemented in an array. The first elements in the array are sorted in ascending order according to their keys, to allow fast search of element in the linked list. We will refer elements in the chunk by their $orderIndex$ (their index in the array). Each element in the list has the following properties:

*orderIndex.next* the order index of the next element in the linked list. Edited only with CAS.

*orderIndex.key* the key of the item. Immutable.

*orderIndex.version* the version of the item. Higher means newer. Edited only with CAS. Initialized with $NONE$, later set to $-GV$ (minus global version), last after inserting the item into the list, set to positive: $version = abs(version)$.
Might get a special value $FREEZE$ by a rebalancing thread.

*orderIndex.dataIndex* the index to the *dataArray* explained below. May be set a negative value to indicate *null* value - which means the key does not exist in the map.

- **Data Array** - The array of the values of the chunk's elements.
$dataArray[orderIndex.dataIndex]$ is the value of the item *orderIndex*. Each cell in dataArray is immutable - it is set only once.

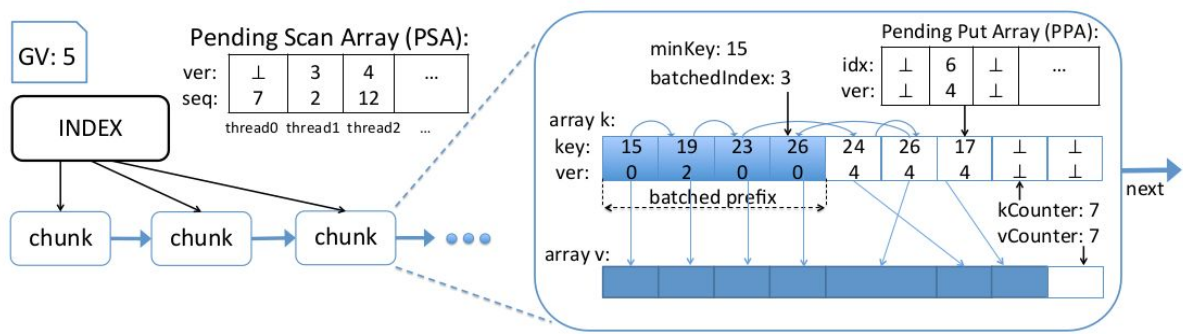

## 1.2.2 Data Structure Invariants

Inside the sorted linked list of each chunk, there is at most one item for each <key, version> pair. Moreover, the items are sorted lexicographically by <key, version>.

Elements are never removed from a chunk's linked list. To discard an element from the map, one puts a null value.

The following variables are changed only with CAS operations, which define a useful synchronization order: item version, item next, item data index.

## 1.2.3 Brief description of basic operations

We bring the essence of the algorithm, while focusing only on details which are relevant to our work.

**Put(key, value)**
1. A put operation begins by allocating place in the arrays: orderArray and dataArray. The allocation is done with an atomic counter increment[1]. After allocating place, it initializes the item in the orderArray: sets version=NONE, key, and writes the data in dataArray.

---

[1] Originally, KiWi allocates place in the dataArray and in the orderArray separately, but this is not required so modified it.

2. Then, it publishes the operation: Writes the orderIndex in the PPA and applies a store fence, $PPA[threadID] = orderIndex$. This publishing allows other threads to see this pending put and help it.
[As long as version=NONE, the item is not considered to be in the map.]

3. Then, it assigns itself a version - by reading the GV, and CAS on the version.

4. After obtaining a version, put adds the item into the linked list - for that purpose, we first apply a fast binary search over the ordered part of the list and then iterate the list until the correct location according to <key, version>.

5. Then, using standard OCC techniques (RETRY on failure), one inserts the element into the linked list. Unless the <key, version> already exists in the list. In that case, one does not insert the item to the list but instead overwrites the dataIndex (leaving the dataArray untouched). One more thing to note, overwriting the dataIndex is not always performed. In case of equality of <key, version> dataIndex defines which item is newer. So we overwrite the dataIndex of the element only if its previous dataIndex is smaller than ours. This overwrite is performed with CAS as well.

6. Last, undo the publishing - $PPA[threadId] = null$

In order to discard an element, one simply invokes $put(key, null)$.
Also, a put might decide to rebalance a chunk and RETRY.

**Helping Put define a version.** Get() and scan() operations iterate the PPA and help each relevant (key relevant to the query) pending put (with version=NONE) to obtain a version. Then, they consider the put as if it was done. They don't help actually inserting it into the linked list. (Before reading the PPA, we apply a fence.)

**Helping Put insert.** When the orderArray of some chunk fills up, one needs to rebalance the data structure. This is done by freezing the chunk, freezing any pending put operations with version=NONE, helping pending puts with negative version and then copying the data (while compacting it on the fly) to new chunks. A negative version means the put is already visible to get() and scan() operations but was not inserted into the linked list yet.

**Get(key)** First, help any relevant put recieve a version. Then, find the relevant chunk. Then find the newest item in that chunk with the appropriate key, while considering also any pending put that was found beforehand in the PPA.

**Scan(minKey, maxKey)** First, increase GV. The scan considers only items with version $\leq$ previous GV. Then, help each relevant pending put recieve a version. Then, find the relevant chunk and scan the chunk's linked list to find all relevant items. For each item, consider only the newest version that is smaller than the scan's version, while newest is defined lexicographically by <version, dataIndex>. While scanning the linked list, consider also relevant items extracted from the PPA.

# 2. Our contribution

## 2.1 Linearizable size bounds

It is useful to determine the map size. However, designing an accurate linearizable size operation for KiWi might be either complicated, inefficient or prone to contention. Such a naive implementation for example is to scan the whole data structure. Therefore, we relax the problem: We define two new operations, that aim to retrieve bounds on the map's size. These bounds provide a good approximation for the map's "real" size.

This is a classic case of tradeoff between complicated interface and performance benefits.

We define two linearizable operations:

sizeLowerBound() - lower bound for the amount of different keys in the map[2] (size).

sizeUpperBound() - upper bound for the amount of different keys in the map (size).

## 2.1.1 Size Lower Bound Design

### 2.1.1.1 The basic idea

We build upon a concurrent adder - LongAdder of Java 8. Assume it is an integer variable with the following linearizable operations: increment(), decrement(), getValue(). Note that in practice LongAdder is not linearizable.

We maintain a single LongAdder for the whole map, $lowerBoundAdder$.

*The basic idea*: **Before** actually removing an element from the map, decrement the lower bound. Later, if we can tell that we didn't actually removed a key - increase the counter. Ideally, it looks like this:

$remove(key)\{$

$\quad\quad lowerBoundAdder --$

$\quad\quad ...$

$\quad\quad if(certainly\ not\ actually\ removed)$

$\quad\quad\quad\quad lowerBoundAdder ++$

$\}$

Similarly, **after** inserting a new key to the map, we can increase the lower bound:

$put(key, value)\{$

$\quad\quad ...$

$\quad\quad if(certainly\ added\ a\ new\ key)$

$\quad\quad\quad\quad lowerBoundAdder ++$

$\}$

This general approach may be used in any data structure to bound various counters.

If we could always determine whether a remove operation actually increased/decreased the elements count, this approach gives an exact size when the map is in idle state - no ongoing operations.

In real-life use cases this is usually the case. However, theoretically we might accumulate gap between the real size and the lower bound

### 2.1.1.2 More details and Pseudocode

Remember that a remove operation is a put of <key, null> item.

Here is a simplified pseudocode for the put operation, changes are highlighted:

---

[2] For discussion simplicity, we assume the map always has a well-defined size. Rigorously, one should discuss various valid linearizations of history of an execution.

```
put(key, value){
        findChunk()
        orderIndex = allocate item in the chunk
        if(value == null) lowerBoundAdder --;
        publishToPPA(orderIndex)
        if(not enough place in the chunk and can freeze the item){
                if(value == null) lowerBoundAdder ++;
                rebalance() and RETRY
        }
        assignVersion(orderIndex)
        if(item was freezed by another thread){
                if(value == null) lowerBoundAdder ++;
                rebalance() and RETRY
        }
        addToLinkedListOfChunk(orderIndex)
}
```

The operations *get*() and *scan*() read the PPA and consider any pending put operations. Therefore the decrement, $lowerBoundAdder$ $--$ must happen **before** publishing to the PPA. Moreover, it may be deferred until that point because no other thread considers the put operation before it is published.

On retry, we know that no other thread "saw the item yet". The item's version was freezed using CAS from NONE to FREEZE_VERSION, meaning any other thread that tries to read the item from the PPA synchronizes with this CAS because it performs a CAS itself before reading the item. So we safely increment the bound $lowerBoundAdder$ $++$ and retry.

Next, we need to explain the meaning of *certainly not actually removed*. We notice, that due to our conservative decrement of the bound before the removal actually takes place - the LongAdder $lowerBoundAdder$ considers all pending operations in the various PPAs. Therefore, it is enough to keep track of how the insertion affects the number of key in the linked lists, or more accurately: "the number of different keys with maximal version and non-null value in the **_linked lists_**".

More, since *rebalance*() helps the *put*() to actually insert the element into the linked list, other thread might perform the actual insertion of the item into the linked list.

So, we define that the responsibility for recognizing "*certainly not actually removed*" and "*certainly added a new key*" is owned by the thread that actually performs the successful CAS that inserts the item to the list (or sets dataIndex in the overwrite scenario).

Here is *addToLinkedListOfChunk*() simplified pseudocode, the changes are marked with green:

```
addToLinkedListOfChunk(orderIndex){
        prev, next = findInsertionLocation(orderIndex)
        if(orderIndex.version == next.version){
                oldDataIndex = next.dataIndex
```

$dataIndex\ =\ orderIndex.dataIndex$

$CAS(next.dataIndex\ =\ dataIndexoldDataIndex\ to)$

**updateCountAfterOverwrite**()

$\}else\{$

$insert\ item\ between\ prev\ and\ next$

$RETRY\ on\ failure$

**updateCountAfterInsert**()

$\}$

$\}$

*Side Note:* "The thread that actually performs the successful CAS" might not exist. When an item is inserted with overwrite - it might never be inserted to the list at all because an item with larger $dataIndex$ might overwrite faster. The overwrite algorithm:

$oldDataIndex = next.dataIndex$

$while(dataIndex\ >\ oldDataIndex\ \&\&\ !CAS(next.dataIndex, oldDataIndex,\ orderIndex.dataIndex))\{$

$oldDataIndex = next.dataIndex$

$\}$

After the loop, we can check whether we actually performed the CAS. If we didn't perform it, then we cannot tell whether it happened because some other helper thread (e.g. that performs *rebalance*()) succeeded with the exact same CAS, or because some other thread inserted an item with a higher dataIndex before us, which means that the item is never actually inserted into the list.

Now, for $updateCountAfterOverwrite$() pseudocode:

$updateCountAfterOverwrite$()

1      $if(not\ actually\ performed\ the\ CAS)\ return$;

2      // certainly not actually removed

3      $if(prev.key\ ==\ key\ \&\&\ value\ ==\ null)\ lowerSizeBound ++$

4      $if(prev.key \neq key\ \&\&\ prev.next\ certainly\ did\ not\ change\ between\ find\ and\ CAS)\{$

5          // The two ifs below are not unified for verbosity purposes.

6          $if(value == null\ \&\&\ data(oldDataIndex)\ == null)$

7              $lowerBoundAdder ++$ // certainly not actually removed

8          $else\ if(value \neq null\ \&\&\ data(oldDataIndex)\ == null)$

9              $lowerBoundAdder ++$ // certainly actually added

10      $\}$

If $prev.key\ ==\ key$ then during the CAS, there is a preceding item in the list with the same key. Since higher versions always appear before lower versions, that preceding item has a higher version. So we can safely undo the conservative decrement $lowerSizeBound ++$.

On the other hand, if $key(prev) \neq key$ then we want to distinguish between two scenarios:

1. A higher version for $key$ was inserted before the CAS:

   Then, we need could undo the conservative decrement.

2. A higher version for $key$ wasn't inserted before the CAS:

   Then, we can safely undo the conservative add or recognize that we added an element as done in lines 6-9.

In practice, we can't distinguish the two cases, but we can handle the most common case. We can usually determine that $prev.next$ did not change since we read it. More details will follow on how we validate that *prev.next certainly did not change between find and CAS* .

In implementing $updateCountAfterInsert()$ , we encounter similar problems:

$updateCountAfterInsert()$

$\quad\quad$ $if(key(prev) == key \,\&\&\, value == null)\, lowerSizeBound\,{+}{+}$

$\quad\quad$ *bool1 = next.dataIndex certainly did not change between find and CAS*

$\quad\quad$ $if(key(prev) \neq key\,)\{$

$\quad\quad\quad\quad$ $if(next.key \neq key \,\|\, bool1)\,\{$

$\quad\quad\quad\quad\quad\quad$ $if(previously\ key\ wasn't\ in\ map\,\&\&\, value == null)$

$\quad\quad\quad\quad\quad\quad\quad\quad$ $lowerBoundAdder\,{+}{+}$  // certainly not actually removed

$\quad\quad\quad\quad\quad\quad$ $else\ if(previously\ key\ was\ in\ map\,\&\&\, value \neq null)$

$\quad\quad\quad\quad\quad\quad\quad\quad$ $lowerBoundAdder\,{+}{+}$  // certainly actually added

$\quad\quad\quad\quad$ $\}$

$\quad\quad$ $\}$

We build upon the fact the the CAS that inserts the new item makes sure that no other item is inserted between them. To determine whether $previously\ key\ wasn't\ in\ map$ : if $next$ has larger key then it is true since the list is sorted and it is guaranteed that items are not inserted between $prev$ and $next$ . Otherwise, one can read the data of $next$ - which is guaranteed not to change. (read the data to see whether it is null or not).
Here, similarly to $updateCounterAfterOverwrite()$ , the hard task is to determine whether *next.dataIndex certainly did not change between find and CAS* .

Now, we handle these two problems. In both cases, we read a variable before a CAS operation, ( $prev.next$ , $next.dataIndex$ ) and we want to make sure that it didn't change before we finish the CAS. Notice that these two variables are edited after their initialization only by using a CAS. Therefore, we can simply read them normally before the CAS, and then read them with a read[3] that synchronizes with CAS. So JMM ensures that they indeed did not change.

Analogically, we also maintain a LongAdder for sizeUpperBound. In the implementation we keep them together in the LowerUpperBounds class, that supplies verbose functions names.

Github: updateCounterAfterOverwrite() [lines 890-905](#)
Github: updateCounterAfterInsert() [lines 937-949](#)

## 2.2 Composition of linearizable operations

### 2.2.1 Motivation

We wanted to make the KiWiMap more user-friendly. A big part of the usage of maps in the industry today is assuming the map used is an extension of AbstractMap. So we decide to try extending the AbstractMap<Integer, Integer> class and see how applicable the KiWiMap can be in the real world.

---

[3] To keep the code uniform, we actually read them with another CAS

First, $keySet()$ and $values()$ are straight-forward to implement above KiWi. We extended the $getRange$ (a.k.a scan) function to collect the keys along side with the values. So, $getRange$ from the Integer.MIN_VALUE to Integer.MAX_VALUE returns all the keys and all the values.

$size()$ and $isEmpty()$ are more subtle, they motivate us to compose $sizeUpperBound()$, and $sizeLowerBound()$ creatively.

### 2.2.2 Composing lower&upper bounds to get isEmpty()

Using a simple optimistic approach, sometimes we can get the answer.

$isEmpty()\{$

   $if(sizeLowerBound() \geq 1)$

      $return\ false$

   $if(sizeUpperBound() \leq 0)$

      $return\ true$

   $return\ unkown$

$\}$

### 2.2.3 Composing lower&upper bounds to get size()

$size()\{$

   $lower1 = sizeLowerBound()$

   $upper = sizeUpperBound()$

   $if(lower1 \geq upper)\{$

      $return\ upper$

   $\}else\{$

      $lower2\ =\ sizeLowerBound()$

      $if(lower2 \geq upper)$

         $return\ lower2$

   $\}$

   $return\ unkown$

$\}$

Assume that the size increments or decrements in single items.

Consider an execution that gives $lower1 \geq upper$ with the following history:

   [t0 [t1 $lowerBound() -> lower1$   t2]   [t3 $upperBound() -> upper$   t4] t5]

We need to show that $upper$ is a valid output for $size()$.

Consider a linearization of the history above. Think of the linearization as if each operation is invoked atomically in a certain time between its $start$ and $end$.

We know that $size \geq lower1$ at some time between $t1$ and $t2$.

We know that $size \leq upper$ at some time between $t3$ and $t4$.

Moreover, $lower1 \geq upper$. Therefore at some point between $t2$ and $t3$,
we have $size = x$ for any $x \in [upper, lower1]$.
So in particular, we have a linearization point between $t2$ and $t3$ with $size = lower1$. Same goes for $lower2 \geq upper$.

Further (non-wait free) heuristics may be applied to increase the chance of getting an answer: for example help ongoing put operations.

Otherwise, backoff. For implementation simplicity, we throw an Exception.

Github: $size()$ [implementation](implementation)
Github: $isEmpty()$ [implementation](implementation)

## 2.3 Performance evaluations

Since we fixed critical bugs in the implementation, the comparison from the paper is irrelevant. We conduct benchmarks similar to the ones run by [1].
We follow the guidelines for reproducible benchmarks from [4].
The keys are always randomized uniformly in the range $\{0, 1, ... MAX\}$, $MAX = 2M$.

### 2.3.1 Competing Data Structures

We benchmark KiWi map against three other maps:
- Java Concurrent Skip-List (Java 8), which does not support linearizable scans.
- The popular Java ConcurrentHashMap (Java 8), which does not support linearizable scans.
- K-ary tree, which supports linearizable scan.

### 2.3.2 Tested Workloads

Every experiment starts with 20 seconds of warmup – inserts, deletes, gets, and scans to a temporary map – to let the HotSpot compiler optimizations take effect.
It then runs 10 iterations, and averages the results and computes their standard deviation. In each iteration, we fill a new map with $INIT\_SIZE$ ($INIT\_SIZE$ is determined below) uniformly randomized <integer, integer> pairs and then exercise some workload for 5 seconds.

We benchmark the data structures on 5 different workloads, with a varying amount of threads:
1. Get-only
2. 50% insert, 50% delete
3. Scan-only - scans of size 32K (scans with keyMin=x, keyMax=x+32K)
4. 50% of the threads perform 50% insert, 50% delete, and the other 50% perform scan-only - scans of size 32K

To make sure the map size is steady[4] in experiments that include insertion/deletion, we use the following rule from [4]:

If your experiment consists of random operations in the proportions $i$% insertions, $d$% deletions, and $s$% *searches* on keys drawn uniformly randomly from a key range of size $r$, then the expected size of the tree in the steady state will be $\frac{r \cdot i}{i+d}$

Both of our experiments that include insertion/deletion - 3 and 4 - have $i = d$ which means $\frac{i}{i+d} = \frac{1}{2}$. Therefore, we have $INIT\_SIZE = \frac{r \cdot i}{i+d} = \frac{1}{2} \cdot MAX = 1M$.

### 2.3.3 Results

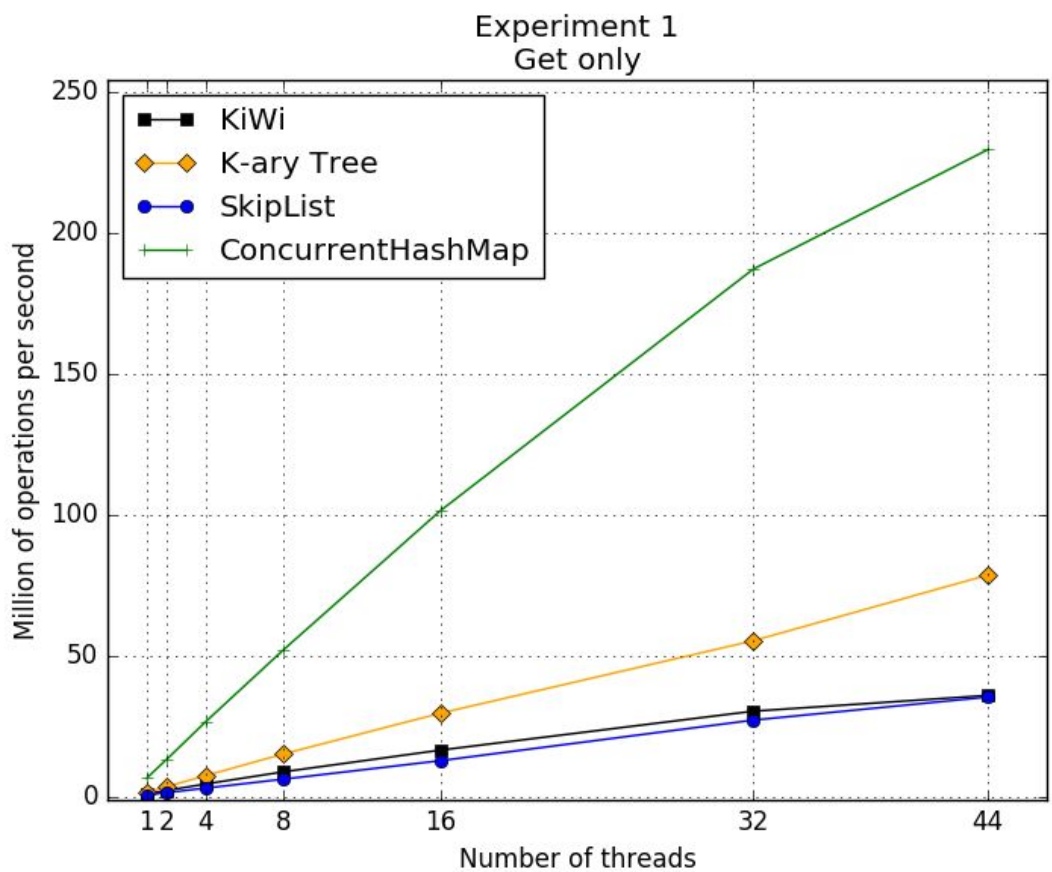

---

4  Note: This keeps the data structure of a steady size, but in our scenario it is not clear that this the right thing to do - because data structures that support linearizable scan inflate with old versions of values.

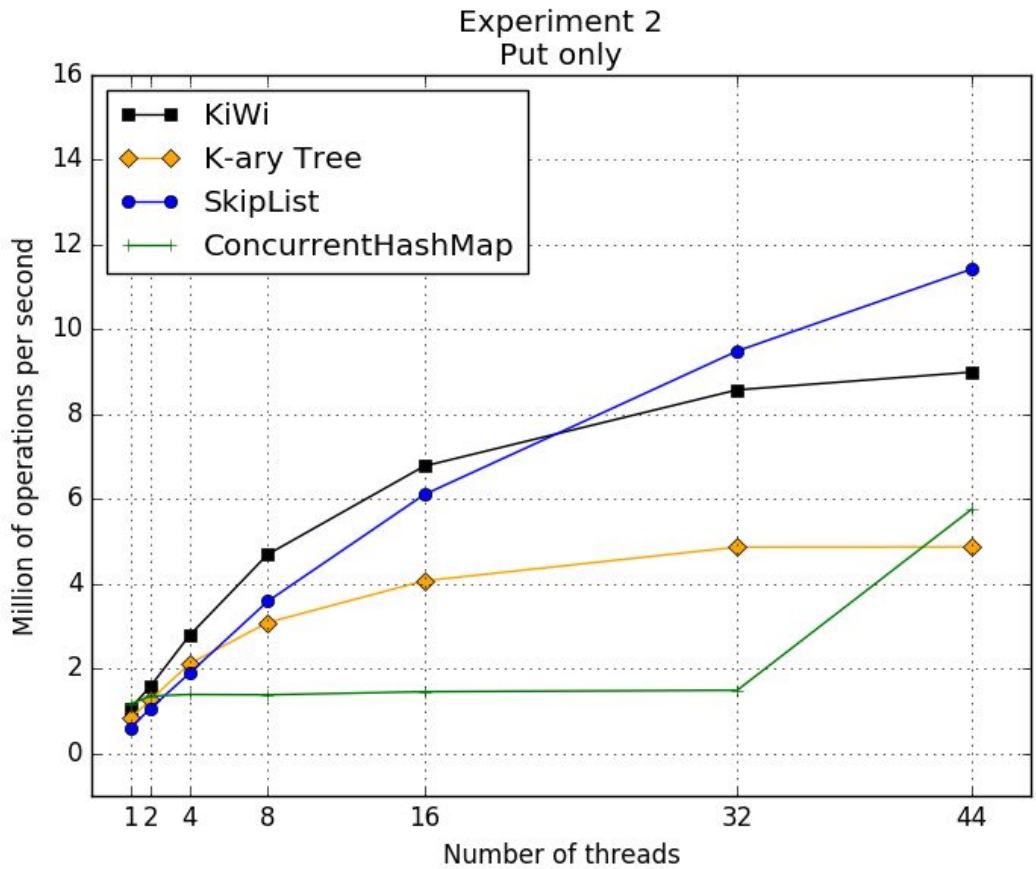

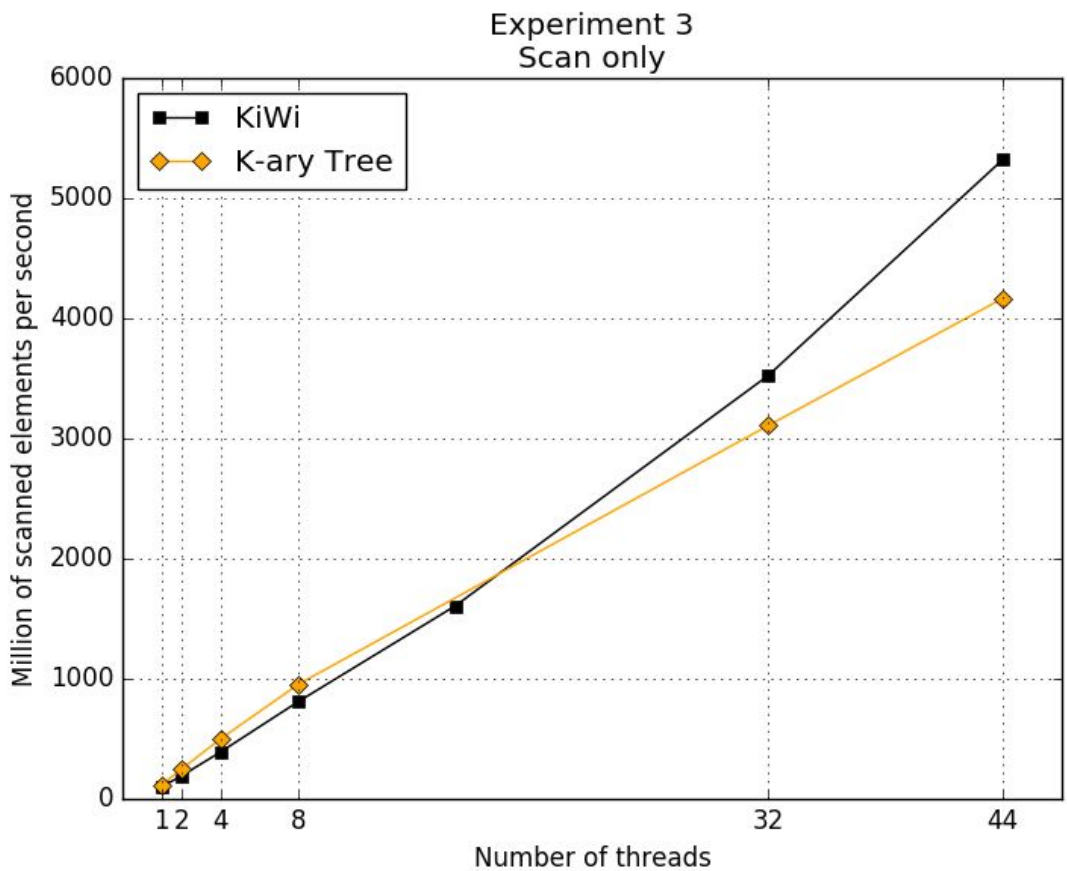

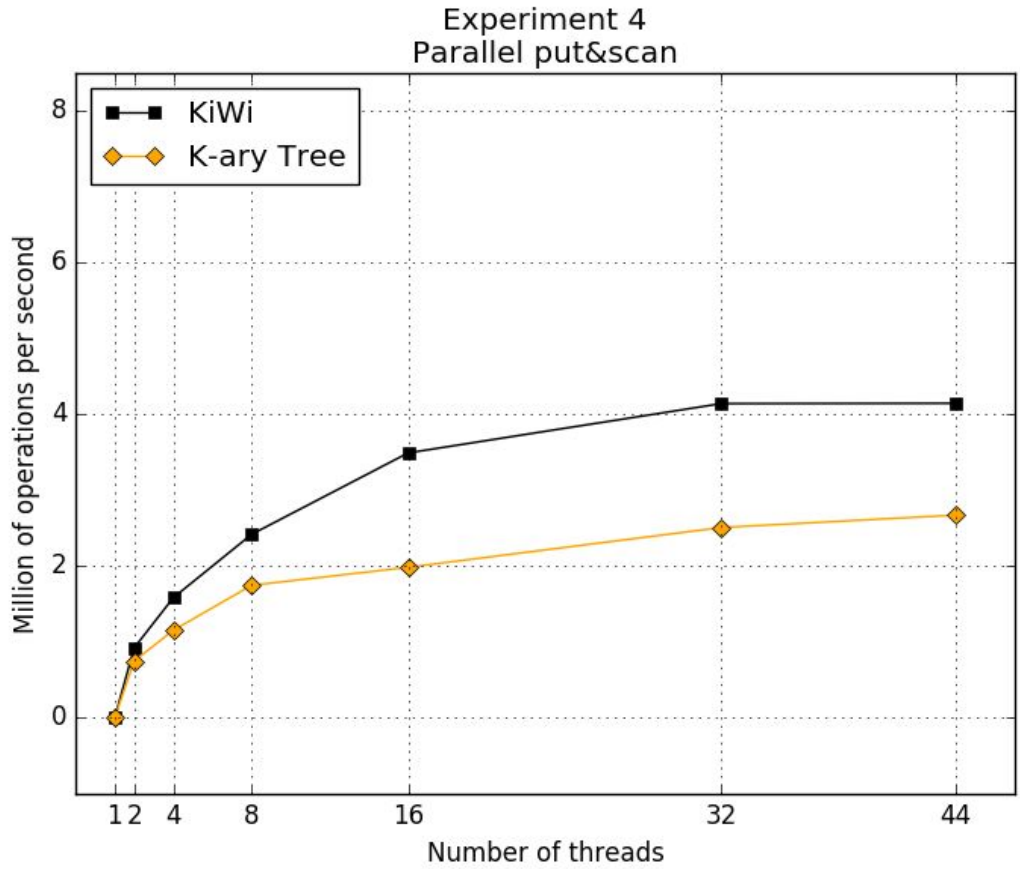

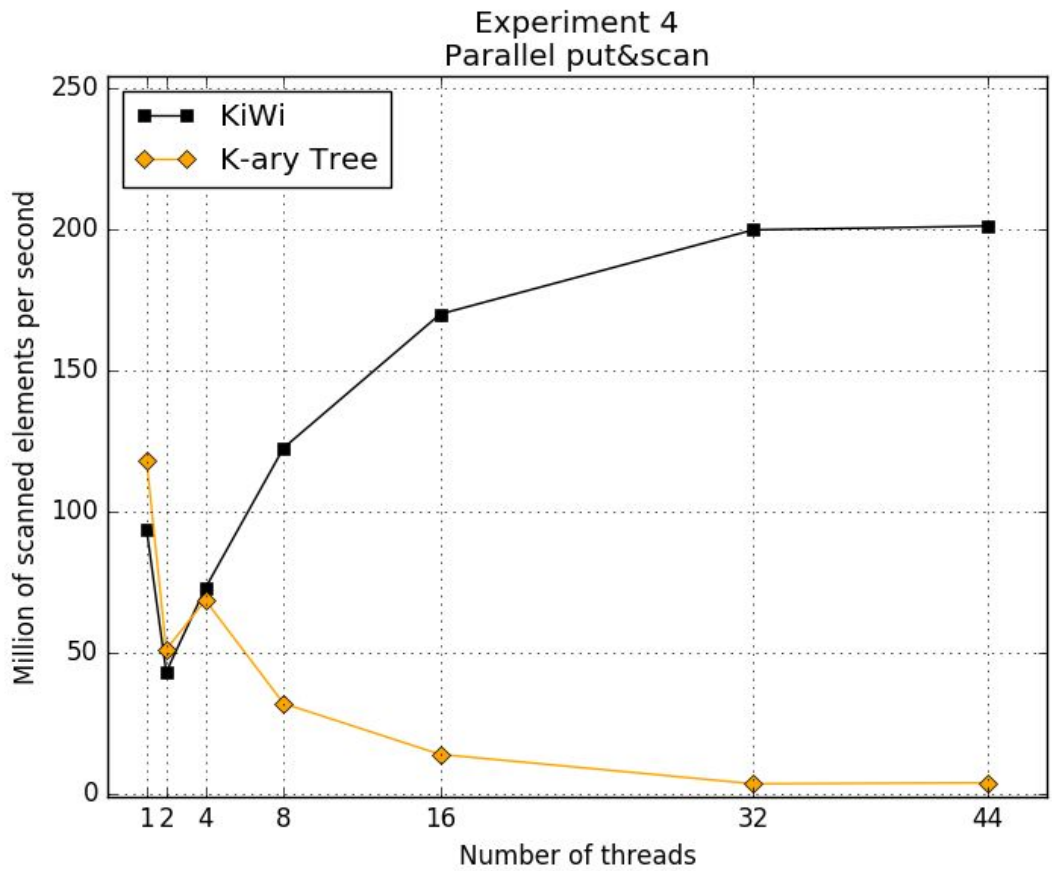

### 2.3.2 Benchmark results discussion

The most significant results are seen on experiments 1  and experiment 4.
We see that KiWi scales well in the put&scan workload, and when talking about get only, the concurrent hashmap works best.
We see that KiWi's advantages still exist after our bug fixes - fast insertions in experiment 2, and scalable scan in experiment 4.
In experiment 4 we experience a phenomena that we couldn't explain in measuring the scan throughput of k-ary tree. Both the measures have a large variance and the average throughput drops down tremendously. Increasing the minimal and maximal heapsize to 20GB did not solve the problem.
Note, that KiWi's implementation is not optimized in terms of performance, also we did not tune the parameters. Therefore, one may hope that it could be much better than other data structures.

### 2.3.4 Technical details

For each measurement, we test the workload 5 times, for 5 seconds each time, ignore the most suspicious result (furthest from the average) and average the remaining 4.
We use the -d64 and -server JVM flags. Also, we allocate just enough memory to the JVM such that garbage collection is performed many times but not too many times. For that purpose, we used 256MB in each experiment.
Each experiment is performed in a separate execution of the JVM.
KiWi's chunk size is set to 4500 (MAX_ITEMS = 4500).
The rebalance policy is tuned as follows: $checkRebalance$ invokes rebalance with probability 0.02 ($rebalanceProbPerc = 2$) whenever the batched prefix consists of less than 0.55 ($sortedRebalanceRatio = 1.8$) of the linked list.
All experiments are performed with the size bounds calculation disabled.
We run Java version 1.8.0_25 on rack-mad-03, (22 cores with 2 hyperthreads each).

To reproduce the results, run the experiments script, and parse them with the parsing script.
Github: experiments script
Github: experiments results
Github: results parsing script

## 2.4 Testing for linearizability

Before starting to edit a piece of code, one must understand what it currently does. This is especially true to an exploratory, partly documented, medium-size piece of code that implements a new concurrent algorithm that is not defined rigorously anywhere and not yet used by anyone.

We generate histories of real executions, and apply a naive implementation of the Wing&Gong [2] linearizability test as described in [3].

More sophisticated linearizability testing methodologies and algorithms exist [3], [5], but we use a naive straight-forward approach and implement it ourselves due to simplicity considerations.

### 2.4.1 Generating histories from real executions

Technically, we randomize operations (insert/discard/scan) run them on a map and log the start time, end time and return value of each operation. That way, we create formal concurrent histories. This is a rather technical project, implemented in the linearizability package.
We implemented each relevant operation, and extended the KiWi map to log the operations it performs, finally we implemented save and load mechanism for such histories, for debugging purposes.

### 2.4.2 Generating useful executions

The first condition, is to generate short histories with only a few threads (preferably 2), because complex histories are cumbersome to debug.
The second condition is to cause any measurement errors become negligible.
The third and last condition is to generate executions which are non-linearizable with high probability. To that purpose, one needs to increase the chance that some subtle change to the data structure is incurred by a thread while another thread is delayed between sensitive opcodes. In practice we inserted random delays between sensitive operations, applying manual fuzzing and educated guesses.

### 2.4.2 Testing a concurrent history

Given a concurrent history, one needs to determine whether it is linearizable. We applied a generic approach, by writing a naive implementation of Wing&Gong naive recursive linearizability test algorithm, as simplified by Lowe [3]. The algorithm is a naive backtracking algorithm over all possible linearization orders of the concurrent history. In each step, it simulates the operation on a standard map and checks the return value.

## 2.5 Bug fixes

Using the linearizability tests, we found a few bugs and fixed them.

### 2.5.1 Comparison by data index and order index

Originally, KiWi defines the newer items lexicographically according to the pair <version, orderIndex>. However, when overwriting an item dataIndex in a put operation the dataIndex is changed while the order index stays the same. This results in new data having a low orderIndex - which caused a bug.
The fix is simply to compare items only according to their dataIndex.

Github: [Line 103](#)

### 2.5.2 Make the get() operation to always consider the item from PPA

In get(), the code considered returning the item found in PPA only if the relevant key existed also in the linked list. This is wrong if the relevant key does not exist in the linked list and the item from the PPA was already seen by some scan() that already finished.

Github: [Lines 784-787](#)

### 2.5.3 Insert null into the linked list even if the key does not exist

When inserting a <key, null> item into the linked list, if the key does not exist in the list - one might think that it may be ignored safely. However, this is wrong because if there is a pending put operation <key, data> with a lower version - we end up with data for the key, which should have been deleted.

Githu: [Remove lines 868-876](#)

### 2.5.4 Rewrite ChunkInt.copyRange() method

The method copyRange is called by a scan() operation. It iterates over a chunks linked list and copies the newest relevant version of each relevant key to the output array. The original implementation tried to apply efficient copyArray() commands, complicated iteration techniques and in practice failed quite simple tests.
The fix is to completely rewrite it using a straightforward implementation.

Github: [Implement the function copyRange()](#)

## 4. Future Work

There are probably a lot more bugs in the implementation and maybe in the data structure non-rigorous design. Our tests focused intra-chunk functionality, while almost ignoring various rebalancing scenarios. Intensive linearizability testing for such scenarios can be applied, by starting with pre-filled chunks, and limiting the chunk size to a smaller amount. More intensive tests can be applied by using a more target specific linearizability testing algorithm. Also, one can apply other testing approaches, such as [5].

We implemented size bounds, but did not measure their performance or accuracy (distance from each other for example). One can repeat the benchmarks while enabling the calculations required for calculating the bounds. One need to define useful measures for the bounds accuracy and measure them experimentally. Then, better size bounds can be designed. The most straight-forward improvement would be to handle more cases when updating the bounds. In cases of uncertainty, a subtle inspection of the orderArray might resolve the uncertainty and shed light on whether one may safely update $sizeLowerBound{+}{+}$ or not.

Maps are commonly filled with big updates, therefore having an efficient putMany may boost the performance. Implementing an efficient linearizable putMany is hard because it might change the whole data structure. A reasonable guarantee definition could be that the items are inserted in a monotonically increasing order, while each put is linearizable. A more efficient putMany that holds this guarantee can be implemented - by inserting consequent items simultaneously.

Originally, we planned to design and implement such an operation but didn't come to it due the need to test the data structure and fix its bugs.

One can tune the parameters of the rebalance strategies to optimize for various workloads/core amount.

One can analyze the chunks' utilization and performance rigorously in simple work loads such insertion/deletion of ascending keys.

Quite simple optimization can be applied, especially in copyRange(), which is the bottleneck of a large scan() operation.

## 5. Acknowledgements

We thank our teacher Adam Morrison for his insightful notes, multi-core machine and for his patience.